# Frustration-induced magnetic bimerons in transition metal halide CoX$_2$ (X = Cl, Br) monolayers

Yu Wang, Shuai Dong and Xiaoyan Yao[*]

Key Laboratory of Quantum Materials and Devices of Ministry of Education,

School of Physics, Southeast University, Nanjing 211189, China

**Abstract**

With the field of two-dimensional (2D) magnetic materials expanding rapidly, noncollinear topological magnetic textures in 2D materials are attracting growing interest recently. As the in-plane counterpart of magnetic skyrmions, magnetic bimerons have the same topological advantages, but are rarely observed in experiments. Employing first-principles calculations and Monte Carlo simulations, we predict that the centrosymmetric transition metal halide CoX$_2$ (X = Cl, Br) monolayers can be promising candidates for observing the frustration-induced bimerons. These bimerons crystallize into stable triangular lattice under an appropriate magnetic field. Compared to the skyrmions driven by the Dzyaloshinskii-Moriya interaction or the long-ranged magnetic dipole-dipole interactions, these frustration-induced bimerons have much smaller size and flexible tunability. Furthermore, the biaxial strain provides an effective method to tune the frustration and thereby to tune the bimeron lattice. In detail, for CoCl$_2$ monolayer, tensile strain can be applied to generate bimeron lattice, further shrink bimeron size and increase the density of bimerons. For CoBr$_2$ monolayer with inherent bimeron lattice state, a unique orientation rotation of bimeron lattice controlled by compressive strain is predicted.

## 1 Introduction

Two-dimensional (2D) van der Waals (vdW) materials in the atomic thickness limit with high tunability and peculiar quantum properties have been significantly highlighted in recent years [1, 2]. Among them, the 2D materials with intrinsic magnetism are particularly intriguing due to plentiful technological opportunities, such as information storage, spintronics and sensing devices [3-5]. Since the first observations of intrinsic ferromagnetism in Cr$_2$Ge$_2$Te$_6$ [6] and CrI$_3$ [7] in 2017, the field of 2D magnetic materials expanded rapidly. The intrinsic magnetism has been confirmed experimentally in a variety of 2D materials, for example, ferromagnetism in Fe$_3$GeTe$_2$ [8] and MnSe$_2$ [9], while antiferromagnetism in FePS$_3$ [10] and CrTe$_2$ [11]. By calculations, more 2D materials have been predicted as promising candidates with intrinsic magnetism, such as GdI$_2$ [12], CrOX (X = Cl, Br) [13], Mn$_3$X$_4$ (X = Te, Se) [14] and so on. Meanwhile, diversified novel phenomena, e.g. multiferroicity, ferrovalley and quantum anomalous Hall effect, have been observed or predicted in these 2D magnetic materials [4, 15-18].

In the field of magnetism, noncollinear magnetic textures with topological untrivial property attracted growing interest for both theoretical and experimental activities [19-21]. In particular, magnetic skyrmions were much highlighted due to its potential applications for nonvolatile energy-efficient spintronic devices [20, 22-24]. It is noteworthy that the skyrmion may show different morphologies under the topological protection, which could be generally called skyrmionic magnetic textures. As an in-plane topological counterpart of skyrmion, the magnetic bimeron has the same advantages, such as topological stability, small size and low driving current density [19, 23, 25]. (Note that bimeron may also refer to "elongated skyrmions" [26-29], which are different from the bimeron in the present paper.) Meanwhile, bimerons are highly desirable due to their unique properties different from skyrmions, for instance, they allow for a pure topological Hall effect upon variation of the in-plane field [19, 30], and they may move without skyrmion Hall effect with a specific current direction [31]. It was first predicted in 2017 that bimeron could exist in an asymmetric form in chiral magnets [32, 33], and also in a symmetric form in the frustrated system [34]. However, compared with the plentiful investigations of skyrmions, bimerons are rarely observed in experiments. Up to now, only isolated bimerons have been observed experimentally [19], for example, the bimeron-like bubbles were observed in single-crystalline Fe/Ni bilayers grown on a W(110) crystal [35], a bimeron was generated in a Py film by local vortex imprinting from a Co disk [36], and recently the antiferromagnetic bimeron was stabilized in α-$Fe_2O_3$ capped with a platinum overlayer [37].

As natural excitations of ferromagnet in 2D space, skyrmionic magnetic textures are much expected to exist in 2D magnetic materials, and the previous experiments demonstrated that the reduction of materials' dimension is beneficial to the stability of skyrmions [38, 39]. It was inspiring that skyrmions had been observed experimentally in the 2D vdW $Cr_2Ge_2Te_6$ and $Fe_3GeTe_2$ [40, 41]. Very recently, room-temperature skyrmion lattice was also reported in experiment of 50% Co–doped $Fe_5GeTe_2$ [42]. On the other hand, different methods were discussed theoretically to generate and modulate skyrmions in 2D magnetic materials [43]. In particular, Janus Cr(I, Cl)$_3$, and polar $VOI_2$ monolayers were predicted to possess Dzyaloshinskii-Moriya interaction (DMI), and thus the bimeron state could be stabilized [44, 45]. Recently, bimerons were also predicted in some vdW heterostructures with DMI [46, 47] and multiferroic $LaBr_2$ bilayer [48].

The frustration of exchange interactions is another important mechanism for skyrmions [49, 50]. In centrosymmetric materials, the frustration-induced skyrmionic magnetic textures may possess more internal degrees of freedom [51, 52]. The stable magnetic textures with different topological charges may coexist, allowing for more advanced logical operations. Compared to the size of several hundred nanometers for most skyrmions driven by DMI or dipole-dipole interaction, the frustration-induced skyrmionic textures are usually much smaller, about several nanometers in size, and thus the higher density may contribute to stronger topological Hall effect [53, 54]. It is pity that the frustration-induced skyrmionic magnetic textures have rarely been reported in 2D materials up to now. The frustration-induced skyrmions and anti-biskyrmions were predicted in vdW $NiI_2$ monolayer [55]. The static and dynamic properties of an isolated bimeron in frustrated ferromagnetic monolayer were studied numerically [56], but the realization of bimerons in real 2D materials remains unsolved and worth more investigation.

In this work, via first-principles calculations and Monte Carlo simulations, we unveil that the frustration-induced bimeron lattice could be realized in 2D centrosymmetric $CoX_2$ (X = Cl, Br) monolayers under appropriate magnetic field. Furthermore, biaxial strain provides an efficient approach to tune the frustration and thus to tune these bimerons. For $CoCl_2$ monolayer, although ferromagnetism dominates, tensile strain can be applied to strengthen frustration and thus induce bimeron lattice. For $CoBr_2$ monolayer with inherent bimeron state, compressive strain may modulate frustration and thus to control the density of bimerons. In particular, it is intriguing that compressive strain can even control the orientation of bimeron lattice in an angle range of 30°.

## 2 Calculation methods

First-principles calculations are performed within the density functional theory (DFT) framework, as implemented in the Vienna *ab initio* Simulation Package (VASP) [57]. The core electrons are treated by the projector augmented wave (PAW) method [58], and the Perdew-Burke-Ernzerhof (PBE) version of the generalized gradient approximation (GGA) is used for structure optimization, electronic structure, and magnetism calculation [59]. The plane-wave cutoff energy is set to be 550 eV. The structures are fully relaxed with the conjugate gradient method until the maximum force is less than 1 meV/Å on each atom, and the total energy converges to $10^{-6}$ eV. For bulk calculation, the vdW correction is described by Many-body dispersion energy method [60, 61]. For monolayer calculations, a vacuum region of approximately 20 Å is added in the perpendicular direction to eliminate spurious interaction between periodic replicas. To sample the Brillouin zone, a k-mesh density of $60/a_0$ is applied along in-plane directions, where $a_0$ denotes in-plane lattice constant. The phonon spectrum is calculated based on the density functional perturbation theory (DFPT) [62, 63]. The *ab initio* molecular dynamics (AIMD) simulation in the canonical ensemble lasts for 3000 fs with a time step of 1 fs at 300 K. The magnetic anisotropy energy (MAE) is calculated with the spin-orbit coupling (SOC) considered.

Furthermore, the Markov-chain Monte Carlo (MC) simulations of the Metropolis algorithm combined with the over-relaxation method are performed on triangular lattice with periodic boundary conditions to study the noncollinear spin textures [64, 65]. The system is first evolved from a relatively high temperature to a very low temperature gradually, and then the energy is further minimized by only accepting the update with the energy reduction to approach zero temperature. The stable magnetic state with the lowest energy is obtained by comparing independent datasets evolving from different initial states on the lattices of different sizes (form 1296 to 5184 sites). And then the bimeron state can be identified from these stable magnetic states by the following analyses.

To characterize the spin ($S$) configuration, the spin structure factor is evaluated for three spin components ($\gamma = x$, $y$ and $z$) respectively as follows:

$$S^{\gamma}(\boldsymbol{q}) = \sum_{i,j} e^{i\boldsymbol{q}\cdot(r_j - r_i)} \left\langle S_i^{\gamma} \cdot S_j^{\gamma} \right\rangle \tag{1}$$

The topological property of magnetic texture is confirmed by topological charge $Q$, namely,

$$Q = \frac{1}{4\pi} \iint \boldsymbol{S} \cdot (\partial_x \boldsymbol{S} \times \partial_y \boldsymbol{S}) dxdy \tag{2}$$

which quantifies the times $S$ wrapping around a unit sphere as the coordinate ($x$, $y$) spans the defined region. If $Q$=1 for each isolated skyrmionic magnetic texture, then the topological charge of whole lattice is just the number of skyrmionic magnetic textures. In this work, the isolated bimerons induced by frustration with topological charge $Q = \pm 1$ are degenerate in energy. To further locate and characterize the topological magnetic textures, the local topological charge density $\rho(\boldsymbol{r})$ is evaluated as follows:

$$\rho(\mathbf{r}) = \frac{1}{4\pi} \mathbf{S} \cdot (\partial_x \mathbf{S} \times \partial_y \mathbf{S}) \tag{3}$$

In simulation, $Q$ and $\rho(\mathbf{r})$ are calculated for a discrete spin lattice in the manner of Ref [66].

## 3 Results

In recent years, transition metal dihalides and trihalides with 2D vdW layered structure attracted considerable attention [67]. Among them, bulk $CoCl_2$ and $CoBr_2$ were synthesized and studied experimentally [68-71]. The stable existence of their monolayers were predicted by later calculations [72-74]. The calculation on basic magnetic properties indicated that for both $CoCl_2$ and $CoBr_2$ the energy of ferromagnetic (FM) order is lower than that of antiferromagnetic (AFM) state, and the magnetism can be controlled via various methods [73, 74]. Since the energy difference between FM and AFM states is relatively small, the more complicated noncollinear magnetic order can be expected.

For the calculation of $CoX_2$ bulks, the Hubbard $U$-correction on the $3d$ orbitals of Co is considered, comparing to the calculation without $U$. It is demonstrated that the in-plane Co-Co distance and the layer spacing without $U$ are more consistent with the experimental data [67-69]. Please see Note 1 in the Supplementary material for details. Note that the previous DFT calculations without $U$ also successfully reproduced the lattice parameters and electronic structures of transition metal halides [72, 75-77]. Therefore, the subsequent results are obtained by the DFT calculation without $U$. It must be clarified that we also perform the calculations of $CoX_2$ monolayers with effective $U = 3$ eV applied, and the results remain qualitatively the same. Please see Note 5 in the Supplementary material for details. Due to the weak interlayer vdW interaction, the cleavage energy ($E_{cl}$) calculated by using four-layer slab method [78] is only about 0.15 J/m$^2$ for both $CoCl_2$ and $CoBr_2$ (Please see Note 2 in the Supplementary material), which is much smaller than that of graphite (0.39±0.02 J/m$^2$) [79]. Thus, it is feasible to exfoliate the $CoCl_2$ and $CoBr_2$ monolayers from their bulks.

$CoX_2$ (X = Cl, Br) monolayer possesses a centrosymmetric crystal structure as shown in Fig. 1(a) and 1(b). From the side view, the monolayer contains Co atomic layer sandwiched by two halogen atomic layers. From the top view, $Co^{2+}$ cations in edge sharing X octahedra construct a triangular net. The relaxed lattice constants $a_0$ of monolayers are summarized in Table 1, consistent well with the previous reports [73, 74].

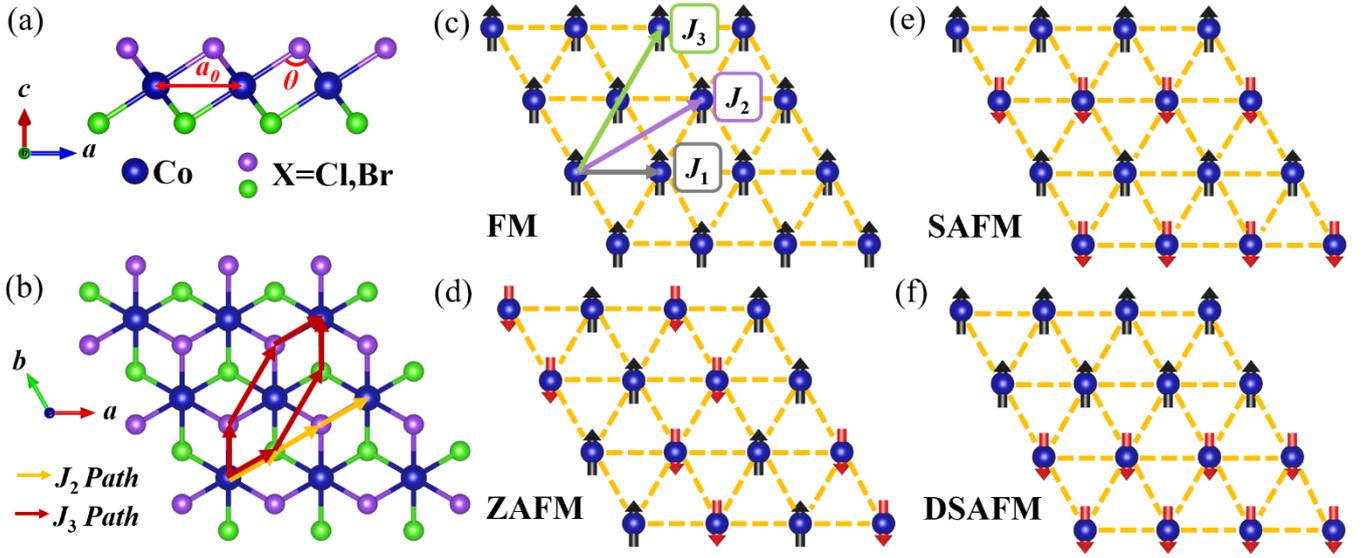

**Fig. 1** The crystal structure of CoX$_2$ (X = Cl, Br) monolayer. (a) The side view where the lattice constant and Co-X-Co bond angle are denoted by $a_0$ and $\theta$, respectively. (b) The top view where Co atoms form an equilateral triangular structure. The X ions in the top and bottom layers are differentiated by green and purple colors. The arrows sketch the possible paths for the second-neighboring and the third-neighboring exchange interactions. (c)-(f) Four basic magnetic orders, namely, (c) Ferromagnetic (FM), (d) Zigzag antiferromagnetic (ZAFM), (e) Stripe antiferromagnetic (SAFM), and (f) Double-stripe antiferromagnetic (DSAFM) configurations.

**Table 1** Calculated lattice constant $a_0$, Co-X-Co bond angle $\theta$, magnetic moment $m$ of Co and X atoms, band gap $E_g$ without and with SOC.

| Units | $a_0$ (Å) | $\theta$ (deg) | $m$ (Co) ($\mu_B$) | $m$ (X) ($\mu_B$) | $E_g$ (eV) | $E_g$ (SOC) (eV) |
|---|---|---|---|---|---|---|
| CoCl$_2$ | 3.545 | 93.240 | 2.517 | 0.173 | 0.369 | 0.366 |
| CoBr$_2$ | 3.748 | 92.680 | 2.460 | 0.188 | 0.373 | 0.400 |

The calculation on the mechanical properties of monolayer CoX$_2$ indicates that the elastic constants meet elastic stability criteria, suggesting their mechanical stability [80, 81]. The calculated Young's moduli are 29.687 N/m for CoCl$_2$ and 28.712 N/m for CoBr$_2$, which are much smaller than that of graphene (340 N/m) with good mechanical flexibility [82, 83]. The gravity-induced out-of-plane deformation estimated by elasticity theory shows that the monolayer CoX$_2$ could withstand its own weight and preserve the planar structure during the process of exfoliation [78, 84]. The dynamical stability is verified by phonon spectra without imaginary frequencies, and the thermal stability is confirmed by AIMD simulation. Please see Note 3 in the Supplementary material for details.

Under the influence of octahedral crystal field, the 3$d$ orbitals of transition metal atom always split into higher energy level $e_g$ ($d_{x^2-y^2}$, $d_{z^2}$) and lower energy level $t_{2g}$ ($d_{xy}$, $d_{xz}$, $d_{yz}$). Thus Co$^{2+}$ ion ($d^7$) has two possible states: high-spin ($S$ = 3/2) and low-spin ($S$ = 1/2). For both CoCl$_2$ and CoBr$_2$, the high-spin state is preferred as shown in Table 1, which has been confirmed by neutron diffraction and previous calculations [71, 85]. As presented in the band structures of Fig. 2, CoCl$_2$ and CoBr$_2$ are both semiconductors, consistent with previous reports [74]. The indirect band gaps are ~0.37 eV for both CoCl$_2$ and CoBr$_2$. It is obvious that Co-$d$ orbitals hybridize with X-$p$ orbitals near the Fermi level. The SOC has weak effect on electronic structures and the band gap only changes a little (Table 1 and Note 4 of the Supplementary material).

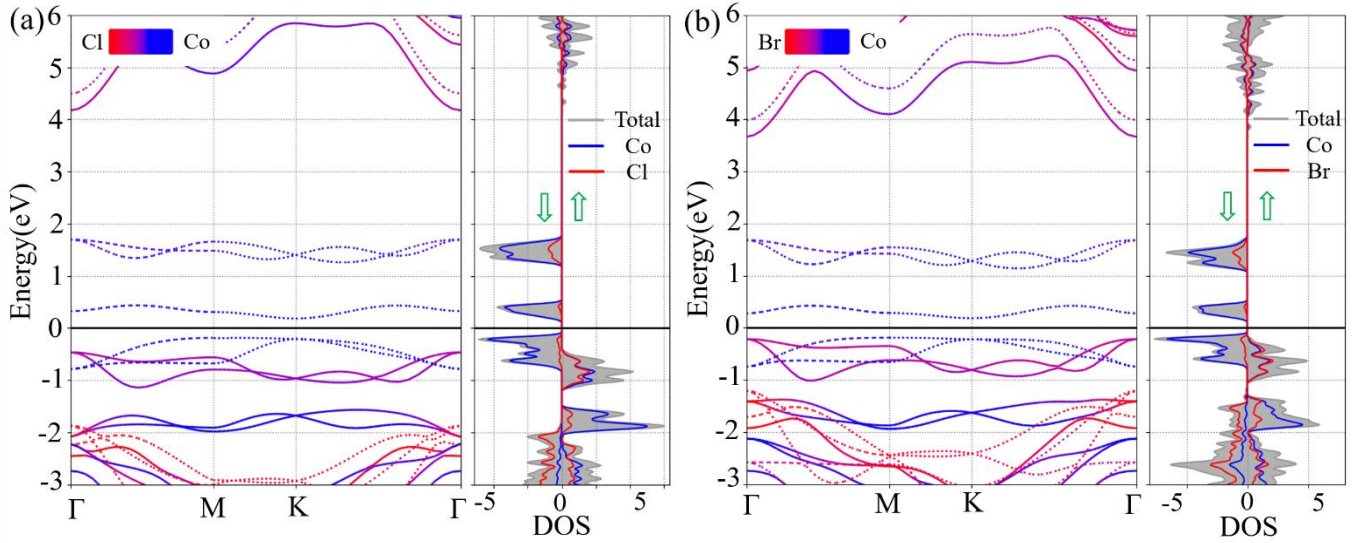

**Fig. 2** Atom-resolved band structures and density of states (DOS) for (a) CoCl$_2$ and (b) CoBr$_2$ monolayers. The solid line represents spin up and the dotted line represents spin down in band structures. The green arrows in DOS indicate spin up and down respectively. The Fermi level is set to 0 eV.

To describe the magnetism of CoX$_2$ monolayer, the classic Heisenberg spin model can be constructed and the Hamiltonian can be written as:

$$H = J_1 \sum_{\langle i,j \rangle} S_i \cdot S_j + J_2 \sum_{\langle\langle i,m \rangle\rangle} S_i \cdot S_m + J_3 \sum_{\langle\langle\langle i,n \rangle\rangle\rangle} S_i \cdot S_n + k \sum_i (S_i^z)^2 + h \cdot \sum_i S_i \quad (4)$$

where $S_i$ represents the normalized classic spin ($|S| = 1$). On the right of Eq. (4), the first three terms describe the exchange energies. $J_1$, $J_2$, and $J_3$ are the nearest-neighboring, second-neighboring, and third-neighboring exchange couplings. The fourth and fifth terms describe the energies of uniaxial anisotropy ($k$: magnetic anisotropic coefficient) and magnetic field ($h$), respectively. Recently, the similar $J_1$-$J_2$-$J_3$ models in the triangular lattice were studied to show AFM skyrmions [86, 87]. In addition, meron/antimeron-like lattice textures were observed in the frustrated honeycomb model with triangular sublattices [88].

To extract the exchange couplings, four basic magnetic orders are considered, namely Ferromagnetic (FM), Zigzag antiferromagnetic (ZAFM), Stripe antiferromagnetic (SAFM) and Double-stripe antiferromagnetic (DSAFM) configurations, as sketched in Fig. 1(c-f). By mapping the DFT energies of these magnetic orders to aforementioned Hamiltonian, $J_1$, $J_2$, and $J_3$ can be obtained as following

$$\begin{aligned} E_{FM} &= E_0 + 3J_1 + 3J_2 + 3J_3 \\ E_{ZAFM} &= E_0 - J_1 + J_2 - J_3 \\ E_{SAFM} &= E_0 - J_1 - J_2 + 3J_3 \\ E_{DSAFM} &= E_0 + J_1 - J_2 - J_3 \end{aligned} \quad (5)$$

where $E_0$ denotes the energy of nonmagnetic part. The calculated values of $J_1$, $J_2$, and $J_3$ are summarized in Table 2. It is indicated that both CoCl$_2$ and CoBr$_2$ show FM $J_1$ ($J_1 < 0$) together with AFM $J_2$ and $J_3$ ($J_2 > 0$, $J_3 > 0$). According to the Goodenough-Kanamori-Anderson (GKA) rules [89-91], the FM $J_1$ mainly derives from the super-exchange interaction between the 3$d$ orbitals of neighbouring Co atoms via overlapping X-$p$ orbitals, where the Co-X-Co bond angle $\theta$ is close to 90° for both CoCl$_2$ and CoBr$_2$ (Table 1). $J_2$ and $J_3$ are both AFM, since they could be considered as super-super-exchange coupling via the $p$ orbitals of two X anions [92]. As discussed in Ref. [92] and sketched in Fig.

1(b), $J_3$ involves two X anions on the same plane, and thus a stronger X-X hybridization, and hence $J_3 > J_2$. CoCl$_2$ has dominant $J_1$ and thus relatively strong ferromagnetism, while CoBr$_2$ possesses $J_3$ comparable to $J_1$, and thus a strong magnetic frustration between exchange interactions. For MAE, the magnetic anisotropic coefficient $k$ is calculated by $E_{oop}$-$E_{ip}$, where $E_{oop}$ and $E_{ip}$ are the FM energies per Co atom with the magnetization along the out-of-plane and in-plane directions respectively. The calculation indicates that CoCl$_2$ has an easy-axis anisotropy ($k < 0$), whereas CoBr$_2$ owns an easy-plane one ($k > 0$) as shown in Table 2. The $k$ of CoBr$_2$ is much larger than that of CoCl$_2$, since the MAE of transition metal halides usually arises predominantly from the SOC in halogen atoms [93].

**Table 2** The exchange coefficients $J_1$, $J_2$, $J_3$ and magnetic anisotropic coefficient $k$

|  | $J_1$ (meV) | $J_2$ (meV) | $J_3$ (meV) | $k$ (meV) |
| --- | --- | --- | --- | --- |
| CoCl$_2$ | -4.188 | 0.073 | 1.229 | -0.015 |
| CoBr$_2$ | -1.906 | 0.110 | 1.910 | 0.575 |

Considering the flexibility of CoCl$_2$ and CoBr$_2$ monolayers, strain provides a powerful approach to tune the magnetic properties. Here, an in-plane biaxial strain is applied, which is described as $\varepsilon = (a-a_0)/a_0$, where $a_0$ and $a$ represent the in-plane lattice constants of pristine and strained monolayers. Aiming at the relatively strong ferromagnetism in CoCl$_2$ and frustration in CoBr$_2$, tensile and compressive strains are applied respectively. In the stain range applied here, $J_1 < 0$ and $J_2, J_3 > 0$ are always kept for both CoCl$_2$ and CoBr$_2$ monolayers.

For CoCl$_2$ monolayer, when the tensile strain ($\varepsilon > 0$) is raised, $J_2$ and $J_3$ decrease slightly, while $J_1$ rises obviously as shown in Fig. 3(a). All the three exchange interactions are weakened as the Co-Co distance is stretched. Meanwhile, the Co-Cl-Co bond angle $\theta$ increases from 93.24° to 96.94°, corresponding to the weakening $J_1$, consistent with the GKA rules [89-91], i.e. the superexchange interaction prefers FM when the cation–anion–cation bond angle is closer to 90°. Although both $J_2$ and $J_3$ are reduced in magnitude by applying tensile strain, the ratio $|(J_2+J_3)/J_1|$ increases as plotted in Fig. 3(b), which ensures the enhancement of frustration. In addition, the magnetic anisotropic coefficient $k$ also changes as the strain is applied, namely the easy direction transfers from out-of-plane to in-plane, as illustrated in Fig. 3(c).

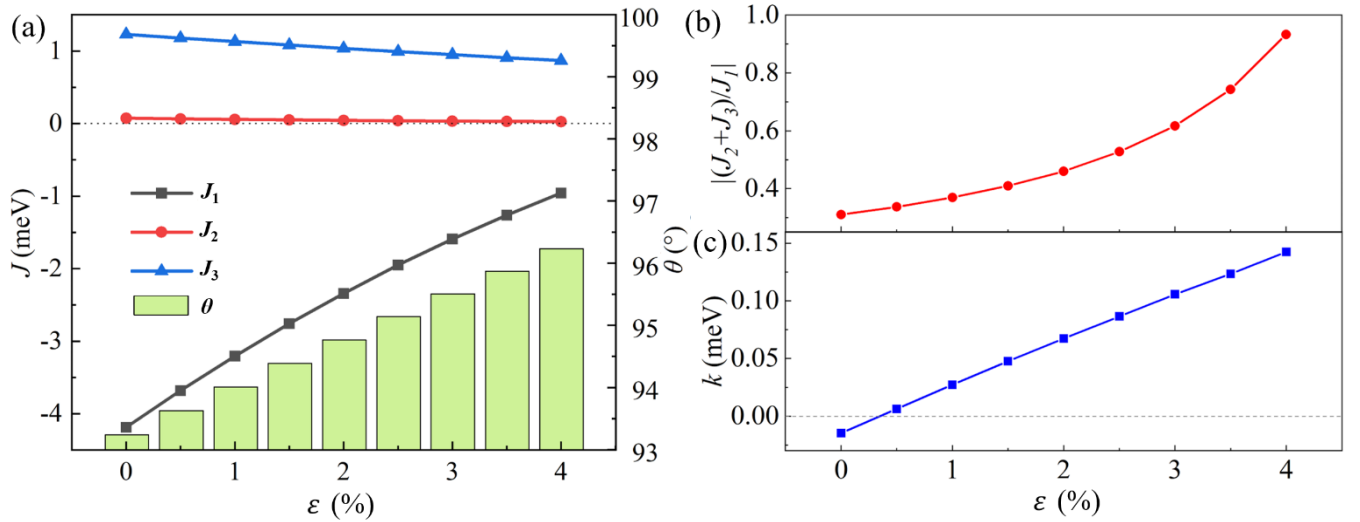

**Fig. 3** The effect of tensile strain on CoCl$_2$ monolayer. (a) The exchange coefficients $J_1$, $J_2$, $J_3$ and Co-Cl-Co bond angle $\theta$, (b) the ratio $|(J_2+J_3)/J_1|$, and (c) the magnetic anisotropic coefficient $k$ as a function of the tensile strain.

In general, the easy-plane anisotropy tends to produce magnetic bimerons [34]. Based on the first-principles parametrized Heisenberg model, the MC simulations are performed. An in-plane magnetic field $\boldsymbol{h}$ is applied along x-axis, because a perpendicular magnetic field is usually required to stabilize skyrmions and a bimeron can be obtained from a skyrmion by 90° rotation around an in-plane axis [30]. The results show that bimerons can be stabilized in $CoCl_2$ monolayer when the tensile strain is applied. As demonstrated in Fig. 4(a), with the strain further increasing, the $h$-window of stable bimeron state with the lowest energy expands, owing to the enhancement of frustration between exchange interactions. Meanwhile, the spiral state exists under the magnetic field lower than that of bimeron region [94].

The real-space spin textures obtained in MC simulation are displayed in Fig. 4(d). Each bimeron is characterized as a meron–antimeron pair with opposite polarities, just corresponding to a region with non-zero local topology charge density $\rho(r)$, as marked in Fig. 4(d). The bimerons form a triangular lattice, corresponding to the typical triple-$q$ spin structure factor with sharp peaks on the corners of a hexagon for $S^y+S^z$, and together with the center for $S^x$, as presented in Fig. 4(b), in contrast to the double-$q$ spin structure factor of meron-antimeron lattices [19, 95]. The strain can effectively modulate the density of bimerons and the size of each bimeron by tuning the frustration. When the tensile strain is raised, the density of bimerons increases gradually [Fig. 4(d)]. Correspondingly, the spin structure factor presents $d_p$ (the distance between the peaks at corner and center) increasing with strain $\varepsilon$. Meanwhile the area of a single bimeron declines, and it reaches as small as about 2.307 nm$^2$ when tensile strain reaches 4%, as shown in Fig. 4(c), which is much smaller than most skyrmions driven by DMI or dipole-dipole interaction. Therefore, the frustration is advantageous in producing more compact noncollinear topological textures, and it can be controlled by strain to tune their density and size, which is in line with the pursuit of higher density and better applicability.

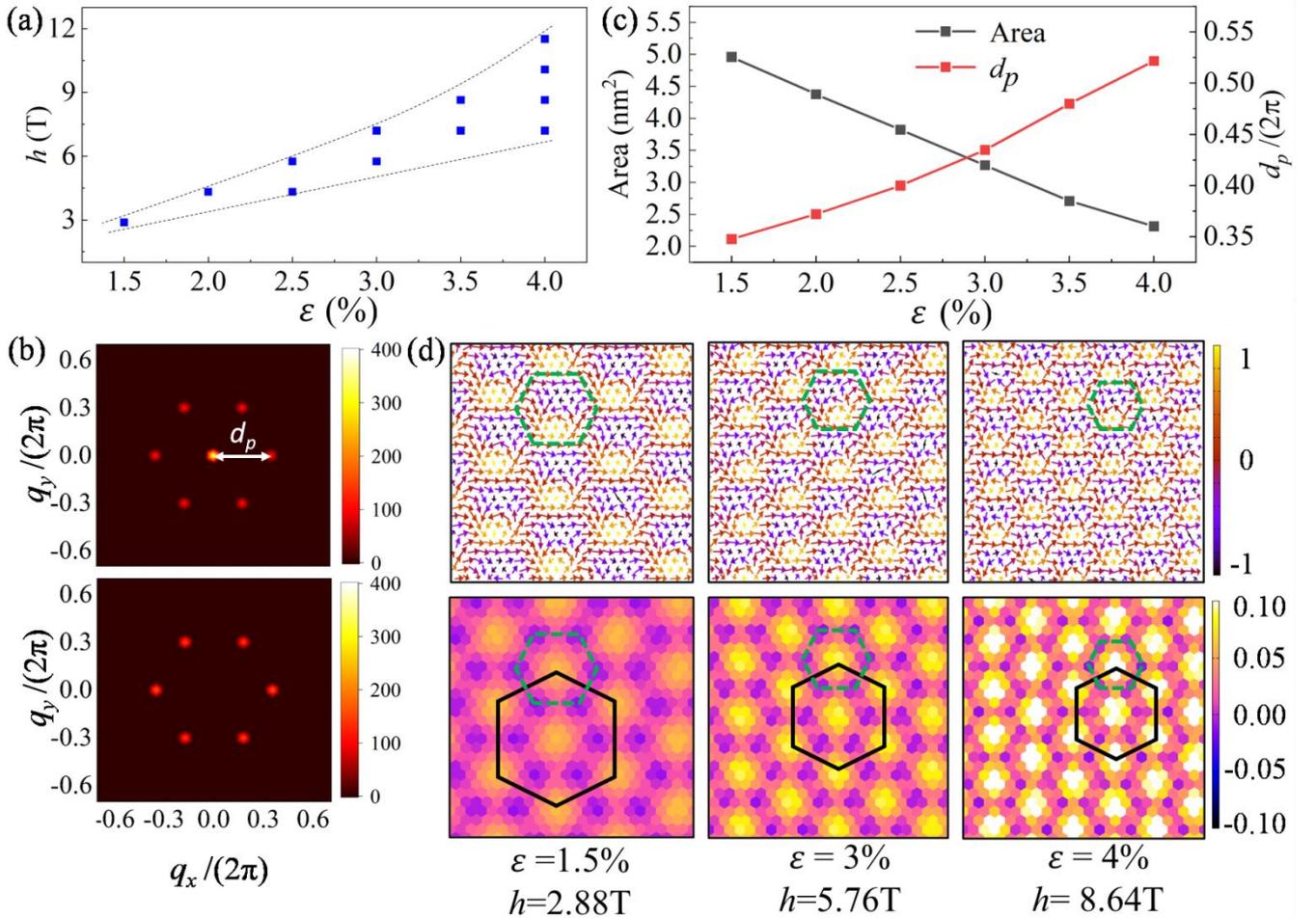

**Fig. 4** The bimeron states in $CoCl_2$ monolayer with tensile strain from MC simulation. (a) The $h$-$\varepsilon$ phase diagram. Here the blue squares

show the parameter points where the bimeron states exist stably. (b) The intensity plots of the spin structure factor at $\varepsilon = 1.5\%$: $S^x(q)$ in the top panel, $S^y(q)+S^z(q)$ in the bottom panel, where $d_p$ denotes the distance between the peaks at corner and center. (c) The area of a single bimeron and $d_p$ as a function of the tensile strain. (d) The real-space bimeron lattices under different tensile strains. The top panels display spin configurations where the vector represents the spin projection onto the *xy* plane and the color presents its *z* component. The bottom panels exhibit the local topology charge density and the color refers to the value of $\rho(r)$. The dashed small hexagon outlines the region of one bimeron, and the solid large hexagon presents the lattice formed by bimerons.

For CoBr$_2$ monolayer, the situation is much more complicated than CoCl$_2$. As plotted in Fig. 5(a), $J_1$ declines first and then ascends. When $\varepsilon$ reaches -3.5%, $J_1$ falls to the minimum (-7.184meV), i.e. the strongest FM value. Meanwhile, the Co-Br-Co bond angle $\theta$ decreases with increasing $\varepsilon$. It reaches 89.92° at $\varepsilon = -3.5\%$, which is extremely close to 90°, just corresponding to the strongest $J_1$, in agreement with the GKA rules [89-91]. Meanwhile, the electronic density between Co and Br ions increases when the strain is strengthened up to -3.5% [inset in Fig. 5(a)], which also implies the enhancement of *p-d* hybridization. On the other hand, it is noteworthy that $J_2$ and $J_3$ exhibit intriguing variations, that is, $J_2$ swaps place with $J_3$. $J_3$ declines from 1.910 to 0.117 meV, while $J_2$ rises from 0.110 to 2.257 meV when $\varepsilon$ is compressed to -4%. Compared to the normal response of $J_2$ and $J_3$ to strain in CoCl$_2$, their subtle variations in CoBr$_2$ may result from the 4*p* orbitals of Br, which are more spatially expanded than 3*p* orbitals of Cl. As shown in Fig. 5(b), although $|J_2/J_1|$ increases but $|J_3/J_1|$ decreases, the ratio $|(J_2+J_3)/J_1|$ declines as a function of $\varepsilon$, which means that the frustration is gradually weakened by compressive strain. In addition, the magnetic anisotropic coefficient *k* of CoBr$_2$ monolayer also decreases with increasing $\varepsilon$, but it always keeps in-plane [Fig. 5(c)].

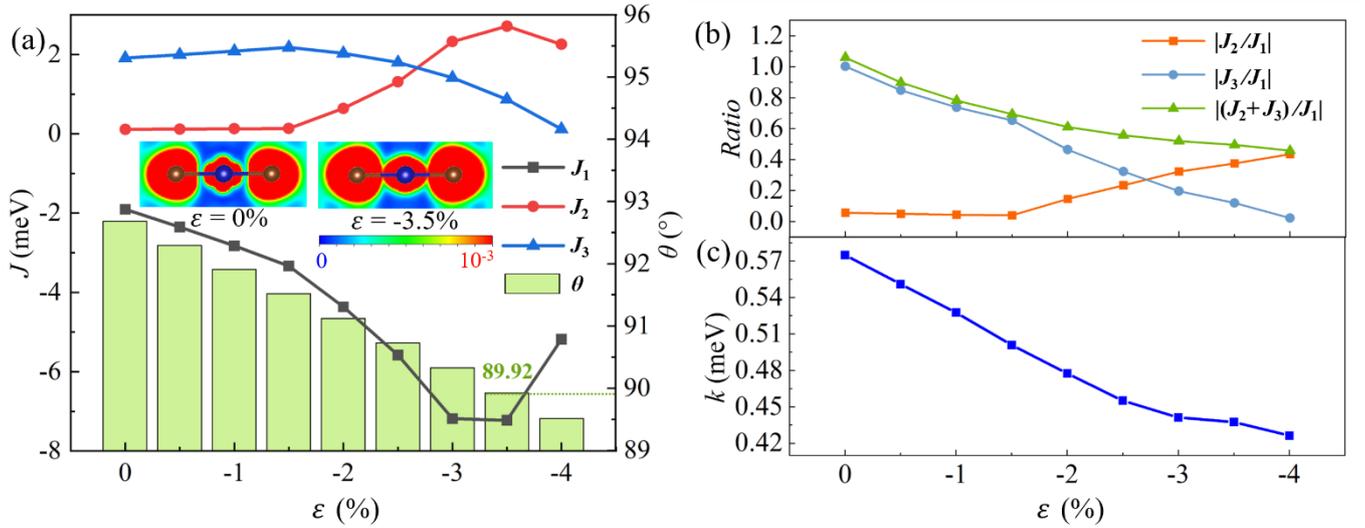

**Fig. 5** The effect of compressive strain on CoBr$_2$ monolayer. (a) The exchange coefficients $J_1$, $J_2$, $J_3$ and Co-Br-Co bond angle $\theta$, (b) The ratios $|J_2/J_1|$, $|J_3/J_1|$, $|(J_2+J_3)/J_1|$, and (c) magnetic anisotropic coefficient *k* as a function of the compressive strain. The insets in (a) present electronic density profiles at $\varepsilon = 0\%$ and -3.5% (cut in the plane of Br-Co-Br bond). The blue and brown balls represent Co and Br atoms, respectively.

The MC simulation of the monolayer CoBr$_2$ under an in-plane magnetic field demonstrates that high-density bimerons can exist in the form of triangular lattice without any strain, as shown in Fig. 6(a) and 6(d). At this time, the area of each bimeron can be as small as 2.281 nm$^2$ [Fig. 6(c)]. These frustration-induced bimerons can be stabilized at finite temperatures [94]. With the compressive strain increasing up to -4%, the bimeron lattice may exist stably under an appropriate magnetic field, but its *h*-window shrinks [Fig. 6(a)]. Meanwhile, the density of bimerons gradually decreases due to the suppression of frustration [Fig. 6(d)]. Correspondingly, the area of a single bimeron expands and the spin structure factor presents $d_p$ decreasing, as presented in Fig. 6(c). It is interesting that when $J_2$ is larger than $J_3$, namely, the strain is stronger than -3%, the orientation of the hexagonal bimeron lattice begins to rotate slightly, as plotted in Fig. 6(d). When the strain reaches -4%, the whole bimeron lattice has successfully rotated 30°. Corresponding to the spin configuration, the peaks of spin structure factor also rotate 30° [Fig. 6(b)]. This rotation of whole bimeron lattice just results from the variation of $J_2$ and $J_3$, namely $J_2$ swaps place with $J_3$ when compressive strain is enhanced from -3% to -4% as shown in Fig. 5(a). Different from the rotation of skyrmion lattice induced by electric currents and thermal gradients in MnSi [96] or by electric field in $Cu_2OSeO_3$ [97], this strain-controlled rotation of bimeron lattice has not been reported, to the best of our knowledge. Therefore, the compressive strain can be applied to control frustration in $CoBr_2$ monolayer to tune not only the density of bimerons but also the orientation of bimeron lattice.

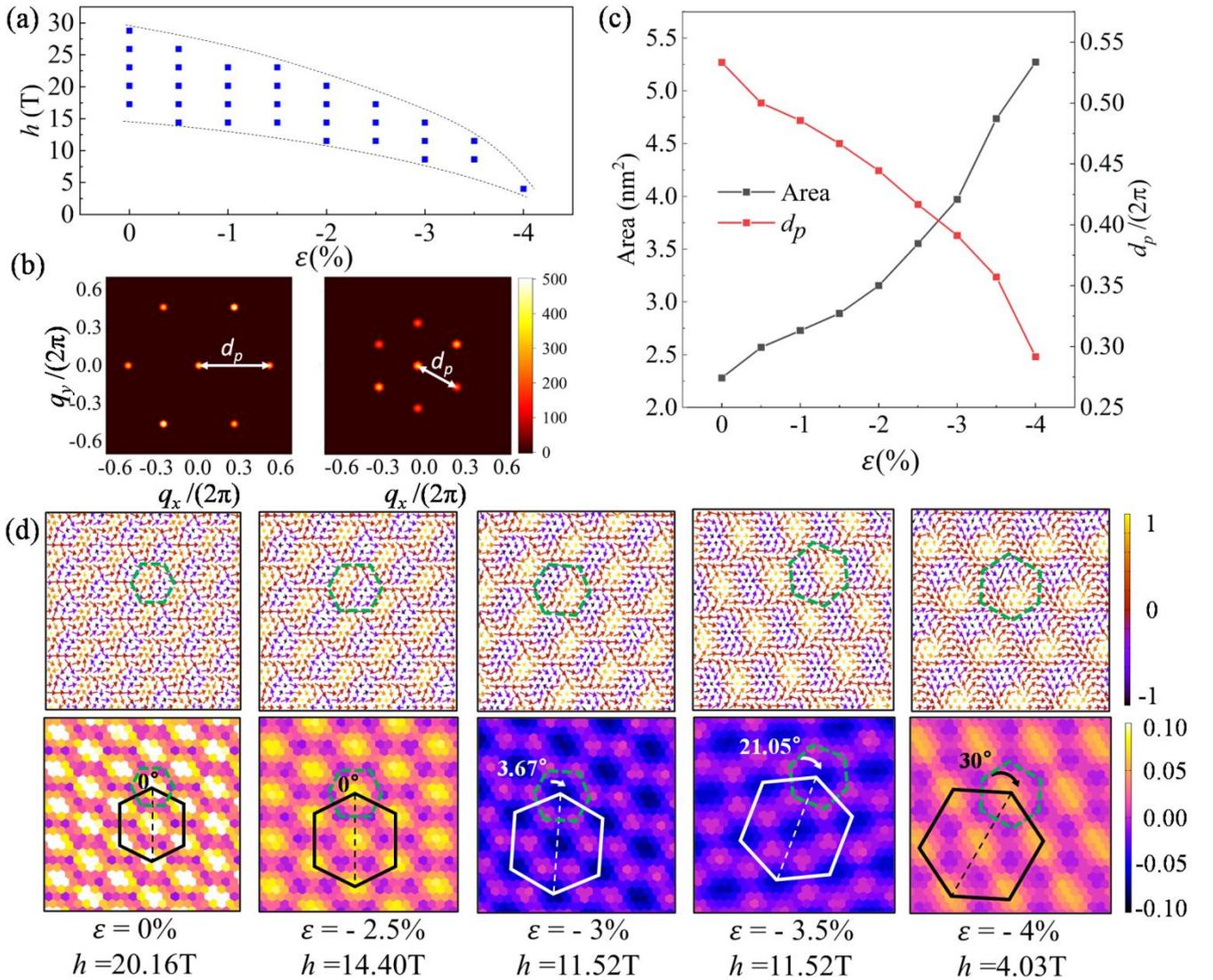

**Fig. 6** The bimeron states in $CoBr_2$ monolayer with compressive strain from MC simulation (a) The $h$-$\varepsilon$ phase diagram. Here the blue squares show the parameter points where the bimeron states exist stably. (b) The intensity plots of the spin structure factor $S^x(q)+S^y(q)+S^z(q)$ for $\varepsilon = 0\%$ in the left panel and $\varepsilon = -4\%$ in the right panel. (c) The area of a single bimeron and $d_p$ as a function of the compressive strain. (d) The real-space bimeron lattices under different compressive strains. The top panels display spin configurations where the vector represents the spin projection onto the $xy$ plane and the color presents its $z$ component. The bottom panels exhibit the local topology charge density and the color refers to the value of $\rho(\boldsymbol{r})$. The frustration-induced bimeron lattices with opposite topological charges are degenerate in energy, so they appear randomly. The dashed small hexagon outlines the region of one bimeron, and the solid large hexagon presents the lattice formed by bimerons.

## 4 Conclusion

In summary, by using first-principles calculations and Monte Carlo simulations, the noncollinear magnetic textures in $CoX_2$ (X = Cl, Br) monolayers are investigated. The calculation indicates that the frustration-induced bimeron lattice could exist under appropriate magnetic field in these 2D monolayers, and biaxial strain provides an effective method to tune these magnetic textures via controlling frustration. For $CoCl_2$ monolayer with dominate ferromagnetism, tensile strain could be applied to generate bimeron lattice. Furthermore, it reduces the size of each bimeron and increases the density of bimeron lattice. For $CoBr_2$ monolayer, bimeron lattice exists even without the application of strain, and it can be modulated by compressive strain. In particular, an exotic strain-controlled rotation of bimerion lattice is revealed in this vdW monolayer. The superiority in size and tunability of these frustration-induced bimerons will trigger special interest to the investigation of the topological magnetic textures in 2D materials, and help for searching more candidates for much-desired spintronic devices with high storage density and low power consumption.

## Acknowledgements

We thank N. Ding, H. P. You and J. Chen for useful discussions. This work is supported by the Natural Science Foundation of Jiangsu Province (BK20221451), the National Natural Science Foundation of China (11834002, 12274070) and the Fundamental Research Funds for the Central Universities (2242022K40027). Most calculations were done on the Big Data Computing Center of Southeast University.